\documentclass[11pt,a4paper,useAMS,usenatbib]{emulateapj}
\usepackage{epsfig}
\usepackage{amsmath}
\usepackage{natbib}

\begin{document}

\title{Detection of a star forming galaxy  \\ in the center of a low-mass galaxy cluster}

\author{\'Akos Bogd\'an\altaffilmark{1}, Lorenzo Lovisari\altaffilmark{1}, Orsolya E. Kov\'acs\altaffilmark{1,2,3}, \\ Felipe Andrade-Santos\altaffilmark{1}, Christine Jones\altaffilmark{1}, William R. Forman\altaffilmark{1}, and Ralph P. Kraft\altaffilmark{1}}
\affil{\altaffilmark{1}Harvard Smithsonian Center for Astrophysics, 60 Garden Street, Cambridge, MA 02138, USA; abogdan@cfa.harvard.edu}
\affil{\altaffilmark{2}Konkoly Observatory, MTA CSFK, H-1121 Budapest, Konkoly Thege M. {\'u}t 15-17, Hungary}
\affil{\altaffilmark{3}E{\"o}tv{\"o}s University, Department of Astronomy, Pf. 32, 1518, Budapest, Hungary}
\shorttitle{SPIRAL GALAXY AT THE CENTER OF A CLUSTER}
\shortauthors{BOGD\'AN ET AL.}

\begin{abstract}
Brightest Cluster Galaxies (BCGs) residing in the centers of galaxy clusters are typically quenched giant ellipticals. A recent study hinted that star-forming galaxies with large disks, so-called superluminous spirals and lenticulars, are the BCGs of a subset of galaxy clusters. Based on the existing optical data it was not possible to constrain whether the superluminous disk galaxies reside at the center of galaxy clusters. In this work, we utilize \textit{XMM-Newton} X-ray observations of five galaxy clusters to map the morphology of the intracluster medium (ICM), characterize the galaxy clusters, determine the position of the cluster center, and measure the offset between the cluster center and the superluminous disk galaxies. We demonstrate that one superluminous lenticular galaxy, 2MASX J10405643-0103584, resides at the center of a low-mass ($M_{\rm 500} = 10^{14} \ \rm{M_{\odot}}$) galaxy cluster. This represents the first conclusive evidence that a superluminous disk galaxy is the central BCG of a galaxy cluster. We speculate that the progenitor of 2MASX J10405643-0103584 was an elliptical galaxy, whose extended disk was re-formed due to the merger of galaxies. We exclude the possibility that the other four superluminous disk galaxies reside at the center of galaxy clusters, as their projected distance from the cluster center is $150-1070$ kpc, which corresponds to $(0.27-1.18)R_{\rm 500}$. We conclude that these clusters host quiescent massive elliptical galaxies at their center. \\
\end{abstract}


\section{Introduction}
\label{sec:intro}

Brightest Cluster Galaxies are luminous and massive galaxies, which are the dominant members of galaxy clusters. In most X-ray--selected galaxy clusters, the BCG is located close to the peak of the diffuse X-ray emission originating from the ICM, implying that BCGs reside at the bottom of the galaxy cluster's potential well  \citep[e.g.][]{hudson10}. 

Observational and theoretical studies demonstrate that BCGs are quenched ellipticals that form through a series of dissipationless mergers \citep[e.g.][]{delucia06,delucia07,collins09,lidman12,lavoie16}. The passive nature of BCGs is maintained by active galactic nuclei, which provide sufficient energy to heat and/or eject the gas from the galaxy, thereby offsetting gas cooling and significant star formation  \citep[e.g.][]{hopkins06,rafferty06,mcnamara07,bykov15}. Therefore, BCGs  do not exhibit significant star-formation rates ($\rm{SFR} \lesssim 1 \ \rm{M_{\odot} \ yr^{-1}}$) with the exception of extreme cooling-flow systems, such as Perseus or the Phoenix Cluster \citep{salome06,mcdonald12,cluver14}. Thus, star-forming disk galaxies are not expected to reside at the center of galaxy clusters. 

In a recent study, \citet{ogle16} analyzed data from the \textit{Wide-field Infrared Survey Explorer} (\textit{WISE}) and the Sloan Digital Sky Survey (SDSS) to explore a population of disk galaxies, called as superluminous spirals and lenticulars. These 53 galaxies are luminous  ($L_{\rm r} = 8-14 \ L^{\star}$), extended ($D=57-134$~kpc), and have high star-formation rates ($5-65 \ \rm{M_{\odot} \ \rm{yr^{-1}}}$). While most of the superluminous disk galaxies reside in relative isolation, eight galaxies may be the BCGs of galaxy groups/clusters. To identify such candidates, \citet{ogle16} searched for known galaxy clusters within $1\arcmin$ of the superluminous disks and compared the redshifts of the galaxies and the galaxy clusters. Although optical observations suggested that superluminous disks are the BCGs, these data could result in misclassification of the BCGs \citep{miller05,linden07}, and, hence it could not be confirmed whether these galaxies reside at the bottom of the galaxy cluster's potential well. Therefore, X-ray observations are essential to  probe whether the superluminous disk galaxies are indeed BCGs and are located at the center of galaxy clusters. 

In this work, we utilize \textit{XMM-Newton} X-ray observations of five galaxy clusters, whose BCGs might be superluminous disk galaxies. We map the morphology of the ICM, determine the position of the X-ray peak, and measure the offsets between the center of the cluster and the position of the candidate BCGs.

\begin{table*}
\caption{Characteristics of the candidate superluminous disk BCGs}
\begin{minipage}{18cm}
\renewcommand{\arraystretch}{1.3}
\centering
\begin{tabular}{ccccccccccc}
\hline
Name &  $z_{\rm gal}$ & $N_{\rm H} $ & $M_{\rm \star}$ & SFR & $kT$  &  $R_{\rm 500}$ &   $L_{\rm 500}$ & \multicolumn{3}{c}{Separation}  \\
           & &($\rm{10^{20} \ cm^{-2}}$) & ($M_{\rm \odot}$) & ($\rm{M_{\rm \odot} \ yr^{-1}}$) &(keV) & (kpc)&($\rm{erg \ s^{-1}}$) & ($\arcsec$) & (kpc)& $1/R_{\rm 500}$ \\
(1) & (2) & (3) &(4) & (5) & (6) & (7) &(8) & (9) & (10)& (11) \\
\hline
J10405643-0103584    & $0.25024$ & $4.5$  &  $4.7\times10^{11}$& 9.3 & $1.90\pm0.40$ & $561$ &  $2.7\times10^{43}$   & $3.928\arcsec$ & $15.5$ & $0.03$ \\
J10100707+3253295   & $0.28990$ & $1.5$  & $5.2\times10^{11}$&25.1 & $4.66\pm0.40$  &  908 & $3.7\times10^{44}$ & $243.840\arcsec$ &  $1067.3 $ & $1.18$ \\
J09260805+2405242  & $0.22239$ &  $3.0$  & $2.6\times10^{11}$&24.0 & $3.71\pm0.57$ &  834 & $3.4\times10^{43}$   & $104.358\arcsec$ & $375.6$ & $0.45$\\
J12005393+4800076   & $0.27841$ &  $2.4$ & $1.2\times10^{11}$& 28.2 & $2.42\pm0.25$ & 630 & $7.7\times10^{43}$  & $123.152\arcsec$ & $523.8$ & $0.83$ \\
J16014061+2718161  & $0.16440$ & $4.0 $   & $3.0\times10^{11}$& 14.8 &  $1.77\pm0.10$ &  564 &$3.7\times10^{43}$  & $53.275\arcsec$ & $150.8$ & $0.27$ \\ 
\hline
\end{tabular} 
\vspace{0.1in}
\end{minipage}
Columns are as follows. (1) 2MASX identifier of the candidate superluminous disk BCG; (2) Redshift of the galaxy from SDSS DR9; (3) Line-of-sight column density based on the LAB survey \citep{kalberla05}; (4) Stellar mass of the galaxy computed from the SED fitting; (5) Star formation rate inferred from \textit{WISE} $12\ \rm{\mu m}$ luminosity \citep{ogle16}; (6) Best-fit gas temperature of the host (or nearby) galaxy cluster;  (7)  $R_{\rm 500}$ radius of the galaxy cluster; (8) $0.1-2.4$ keV band luminosity of the galaxy cluster within the $R_{\rm 500}$ radius obtained from the \textit{XMM-Newton} observations; (9), (10), and (11) The projected distance between the candidate BCG and the center of the galaxy cluster in units of arc second, kpc, and $R_{\rm 500}$ radius, respectively. 
\label{tab:clusters}
\end{table*}

\begin{figure*}[!]
  \begin{center}
    \leavevmode
      \epsfxsize=0.32\textwidth \epsfbox{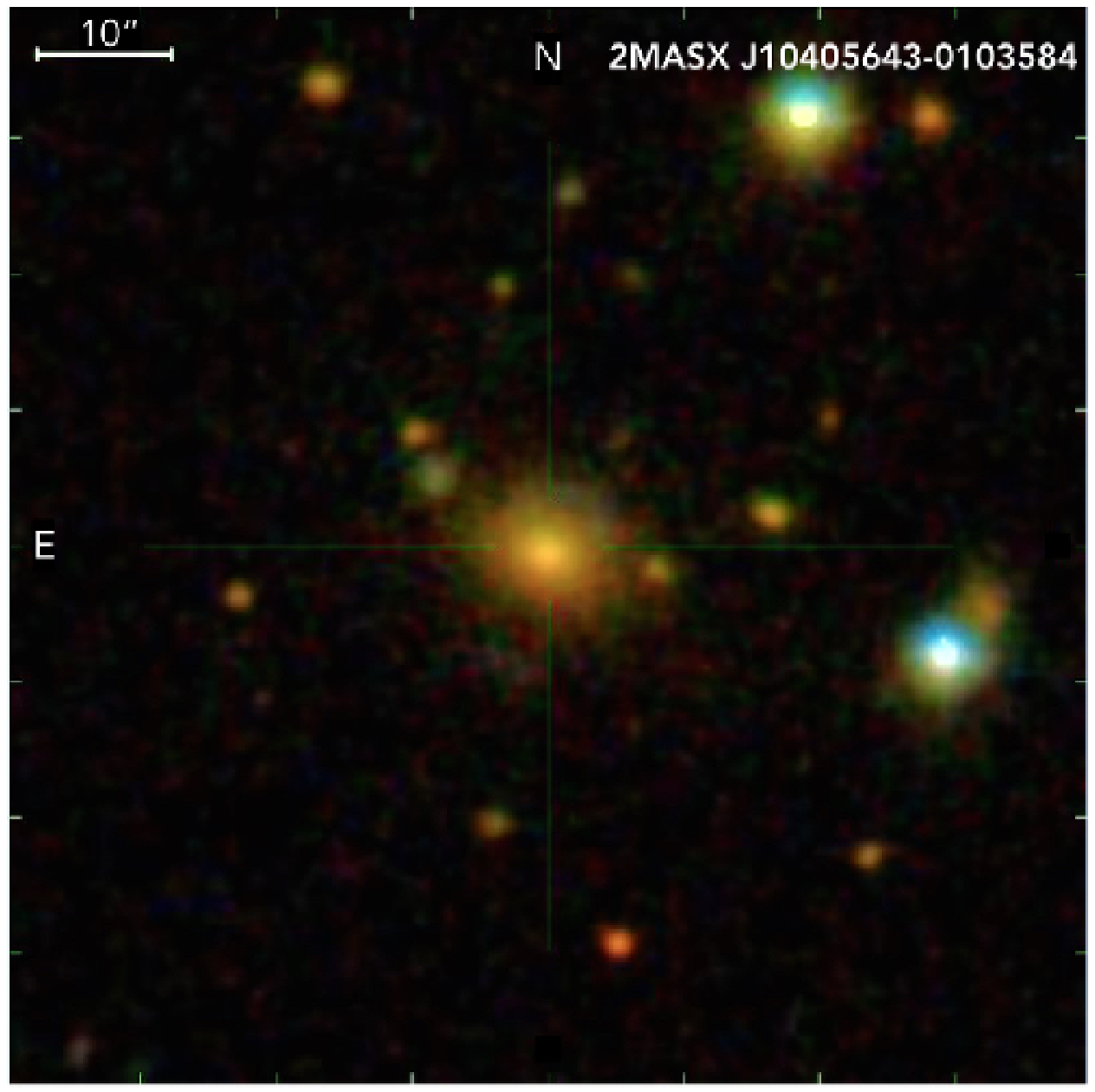}
      \epsfxsize=0.32\textwidth \epsfbox{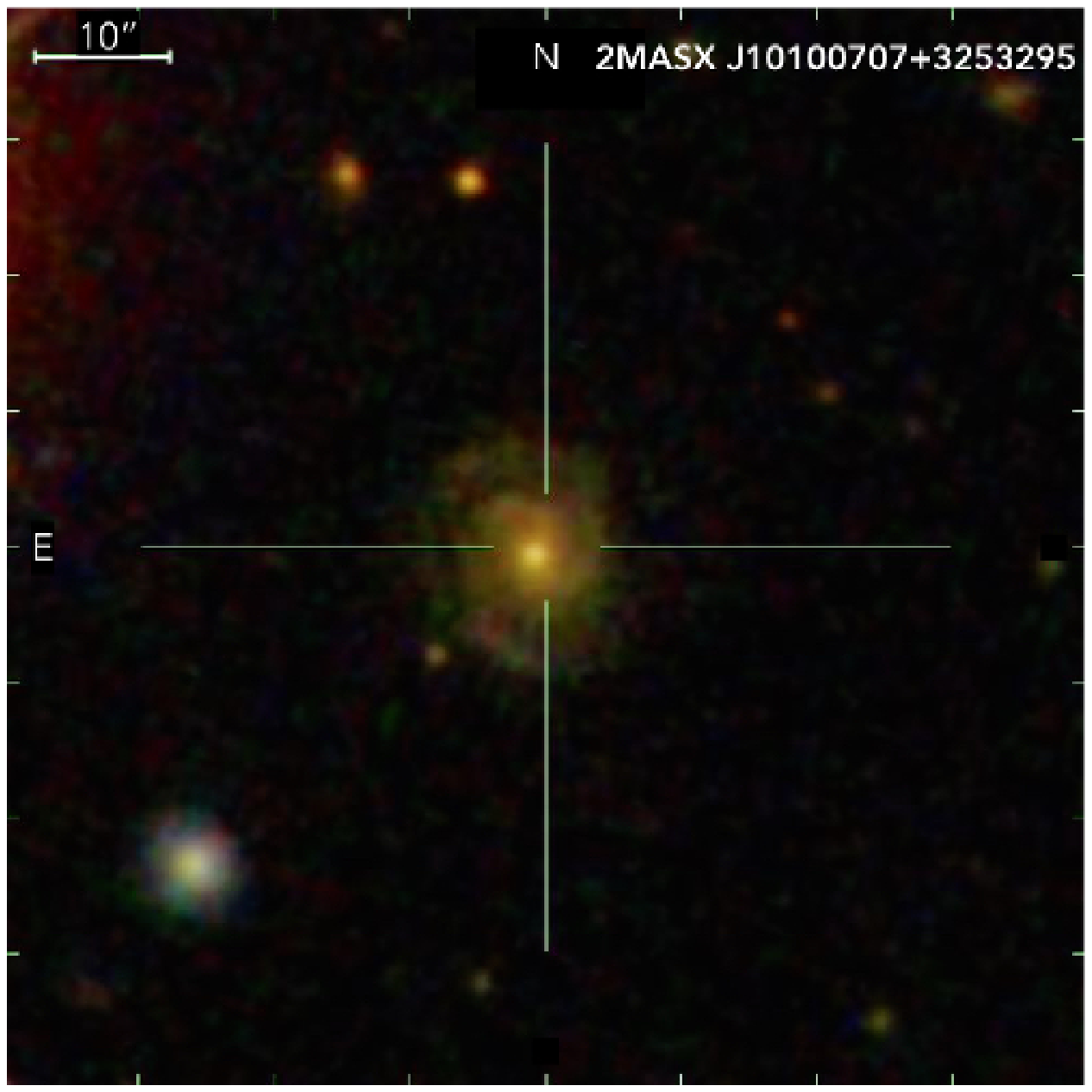}
      \epsfxsize=0.32\textwidth \epsfbox{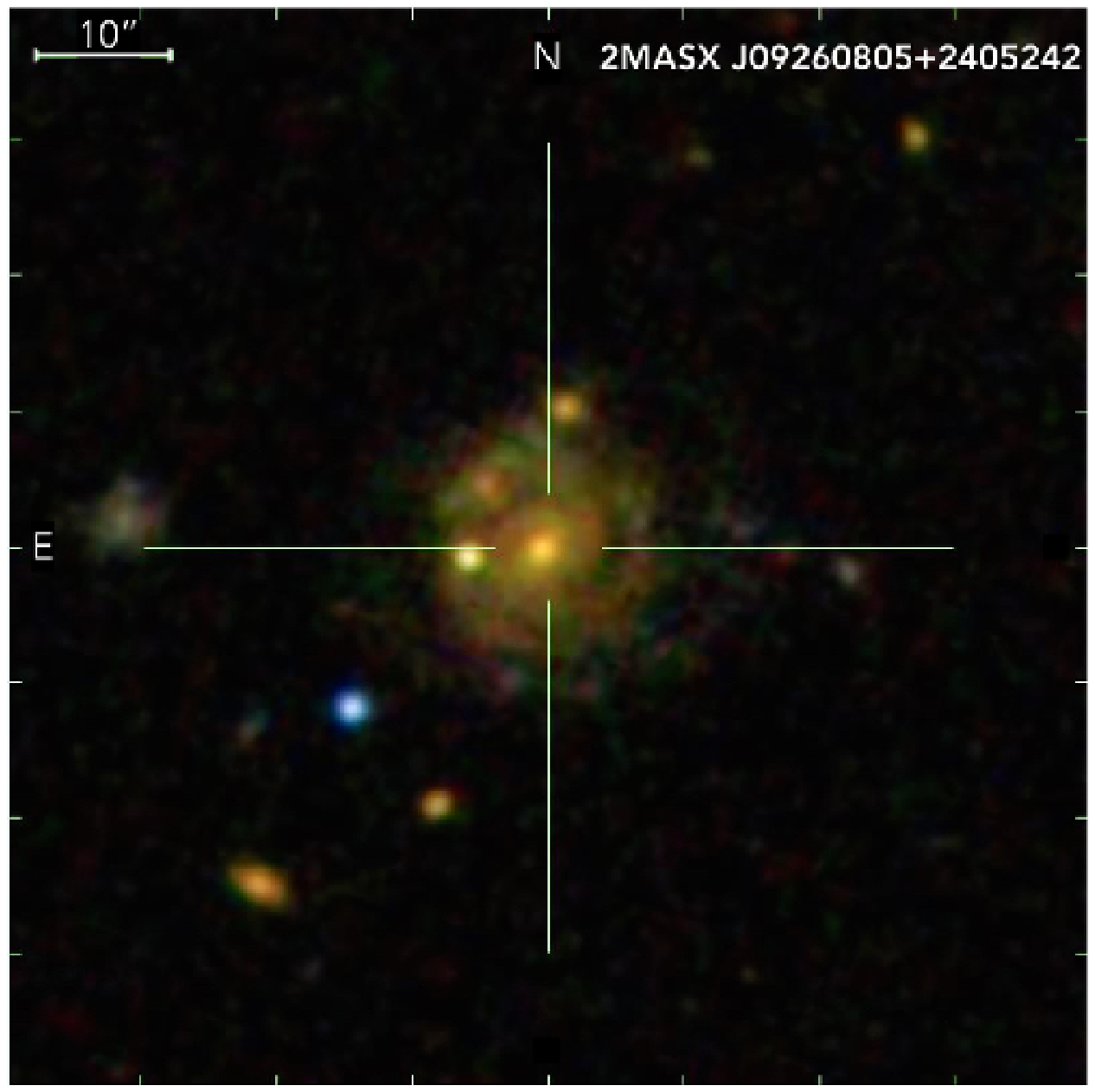}
      \epsfxsize=0.32\textwidth \epsfbox{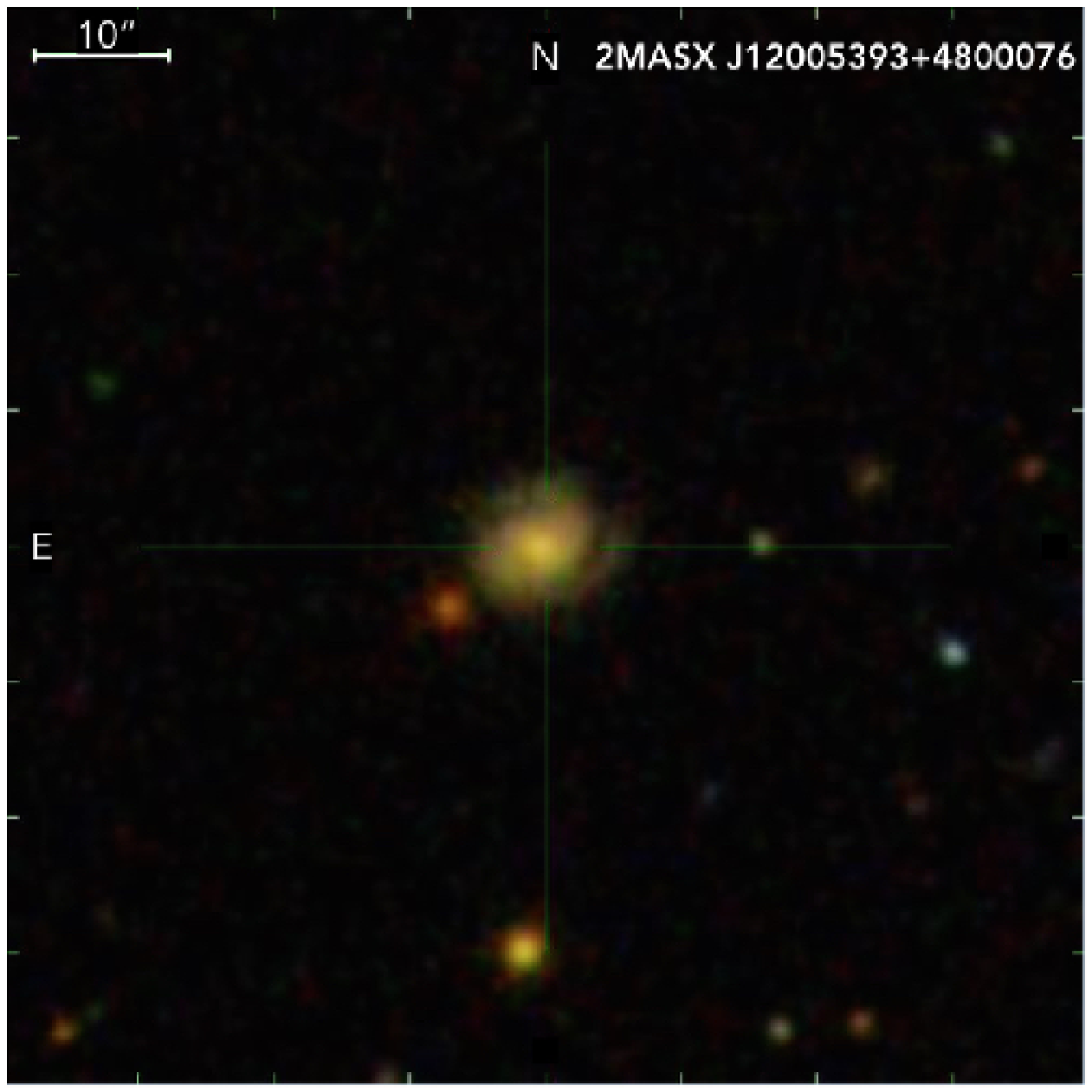}
      \epsfxsize=0.32\textwidth \epsfbox{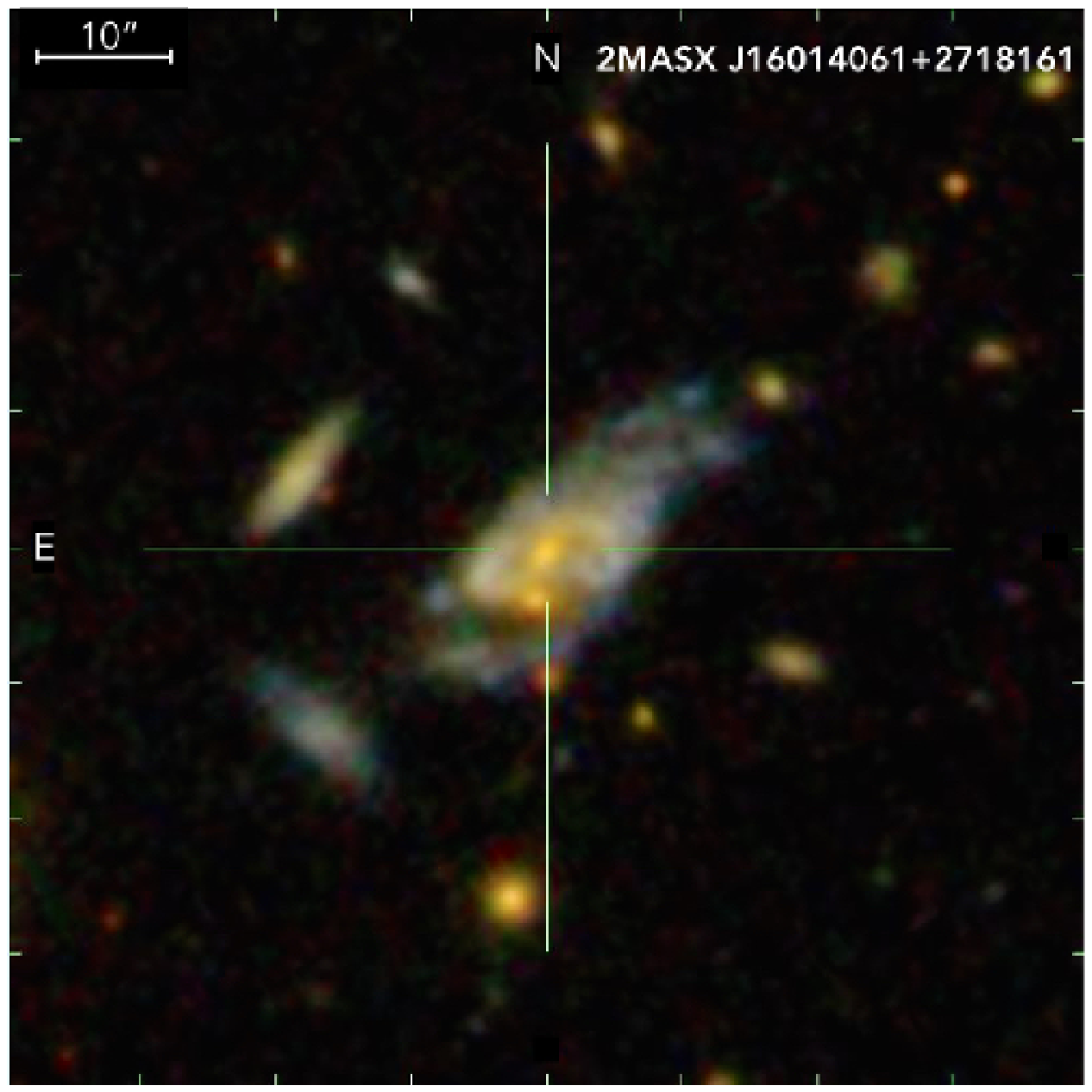}
      \vspace{0cm}
      \caption{SDSS $g$-band images of the five candidate superluminous disk BCGs. The disk galaxies in our sample are massive, $M_{\rm \star} = (1.2-5.2)\times 10^{11} \ \rm{M_{\odot}}$, and exhibit star formation rates in the range of $(9.3-28.2) \ \rm{M_{\odot} \ yr^{-1}}$. The galaxies are called as superluminous spirals and lenticulars due to their large, $63-87$ kpc, diameters, which significantly exceed the size of disk galaxies with similar mass.} 
     \label{fig:candidates}
  \end{center}
\end{figure*}

Throughout the paper we assume $H_0=70 \ \rm{km \ s^{-1} \ Mpc^{-1}}$, $ \Omega_M=0.27$, and $\Omega_{\Lambda}=0.73$, and all error bars are $1\sigma$ uncertainties. \\

\bigskip

\section{The galaxy cluster sample}
\label{sec:sample}

The sample of \citet{ogle16} contains eight spirals and lenticulars that may be BCGs of galaxy clusters. To estimate the luminosity of these galaxy clusters, we first utilized archival \textit{ROSAT} All Sky Survey (RASS) observations.  We detected five galaxy clusters at the  $\gtrsim1.3\sigma-3.3\sigma$ significance level in the \textit{RASS} images, while the galaxy groups/clusters associated with three candidate BCG spirals (2MASX J09260805+2405242,  2MASX J11535621+4923562, CGCG 122Ð067), remain undetected. To estimate the luminosity of the galaxy clusters, we convert the observed count rates to flux assuming that the ICM has a temperature of $kT=2$~keV. We estimate that the five detected galaxy clusters have luminosities $L_{\rm0.1-2.4kev} \gtrsim 3\times10^{43} \ \rm{erg \ s^{-1}}$, while  the three non-detected systems may be fainter X-ray groups with $L_{\rm0.1-2.4kev} \lesssim 3\times 10^{43} \ \rm{erg \ s^{-1}} $. 

Because of the short exposures and the low angular resolution of the \textit{RASS} images, these data are not suitable to probe whether the superluminous spirals and lenticulars reside in the center of the clusters. Therefore, we collected \textit{XMM-Newton} X-ray observations of the five galaxy clusters to determine the positions of the galaxy cluster centers and analyze the characteristics of the clusters. The properties of the candidate BCGs and the associated galaxy clusters  are listed in Table \ref{tab:clusters}.

\begin{table*}
\caption{The list of the analyzed \textit{XMM-Newton} observations}
\begin{minipage}{18cm}
\renewcommand{\arraystretch}{1.3}
\centering
\begin{tabular}{c c c c c }
\hline 
Galaxy & Obs ID & $t_{\rm total}^{\dagger}$  & $t_{\rm clean}^{\ddagger}$ & Date \\
name &  & (ks)  & (ks)& \\
\hline
2MASX J10405643-0103584 & 0654080301 & 16.4 &9.2/9.4/6.9& 2010 Dec 07 \\
2MASX J10100707+3253295 & 0802750101 & 16.0 &6.6/6.7/6.8  & 2017 Oct 24 \\
2MASX J09260805+2405242 & 0802750201 & 17.0 &9.7/10.1/3.7  & 2017 Apr 16  \\
2MASX J12005393+4800076 & 0802750301  & 21.0 &17.1/17.1/12.7  & 2017 May 04\\
2MASX J16014061+2718161 &  0802750501  & 45.9 &29.0/29.6/22.7  & 2017 Sep 08 \\
 \hline \\
\end{tabular} 
\end{minipage}
$^{\dagger}$ Total exposure time of the observations. \\
$^{\ddagger}$ The clean exposure times refer for the EPIC PN, MOS1, and MOS2 cameras, respectively.
\vspace{0.5cm}
\label{tab:xmmdata}
\end{table*}

\section{Analysis of the \textit{XMM-Newton} data}
\label{sec:data}

We utilize \textit{XMM-Newton} X-ray observations to study the five galaxy clusters in our sample. All observations were taken with the European Photon Imaging Camera (EPIC). We analyzed the data using the XMM Science Analysis System (SAS) version 15.0 and Current Calibration Files (CCF). The analyzed list of observations is listed in Table  \ref{tab:xmmdata}. 

The data are analyzed following our earlier works \citep{lovisari15,lovisari17,bogdan18}. We generate the calibrated event files using  \textit{emchain} and \textit{epchain} tasks and include event patterns $0-12$ for EPIC-MOS and 0 for EPIC-PN data.  We identify and exclude  the high background time periods using a two-step filtering approach. We first used the light curves from the $10-12$ keV and $12-14$ keV bands for EPIC-MOS and EPIC-PN and applied $2\sigma$ clipping. Second, we constructed the $0.3-10$ keV band light curve of the previously cleaned event files to remove any residual high background periods. Even after filtering the solar flare events some observations are still affected by soft proton contamination. We estimated that contribution performing the Fin/Fout ratio calculation suggested by \citet{deluca04}. The clean exposure times are listed in Table \ref{tab:xmmdata}. 

To detect bright point sources, we used the \textit{edetect\_chain} tool. The source list was created from the $0.3-10$ keV band images and was used to mask point sources from the analysis. To account for the instrumental and sky background components, we follow the approach described in \citet{lovisari15,lovisari17}.

\begin{figure*}[!t]
 \begin{center}
        \includegraphics[width=0.4\textwidth]{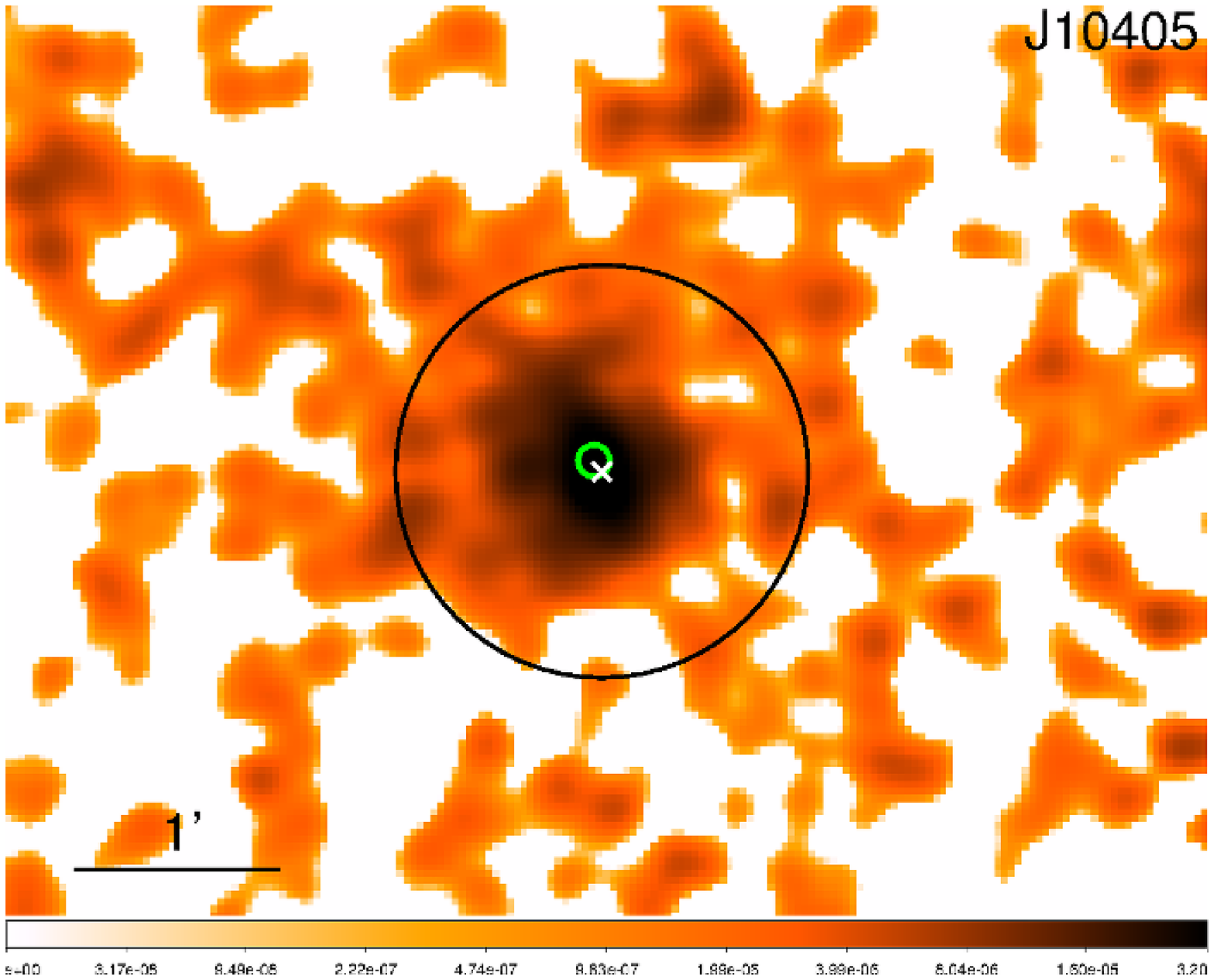} \hspace{0.2cm} \vspace{0.2cm}
        \includegraphics[width=0.4\textwidth]{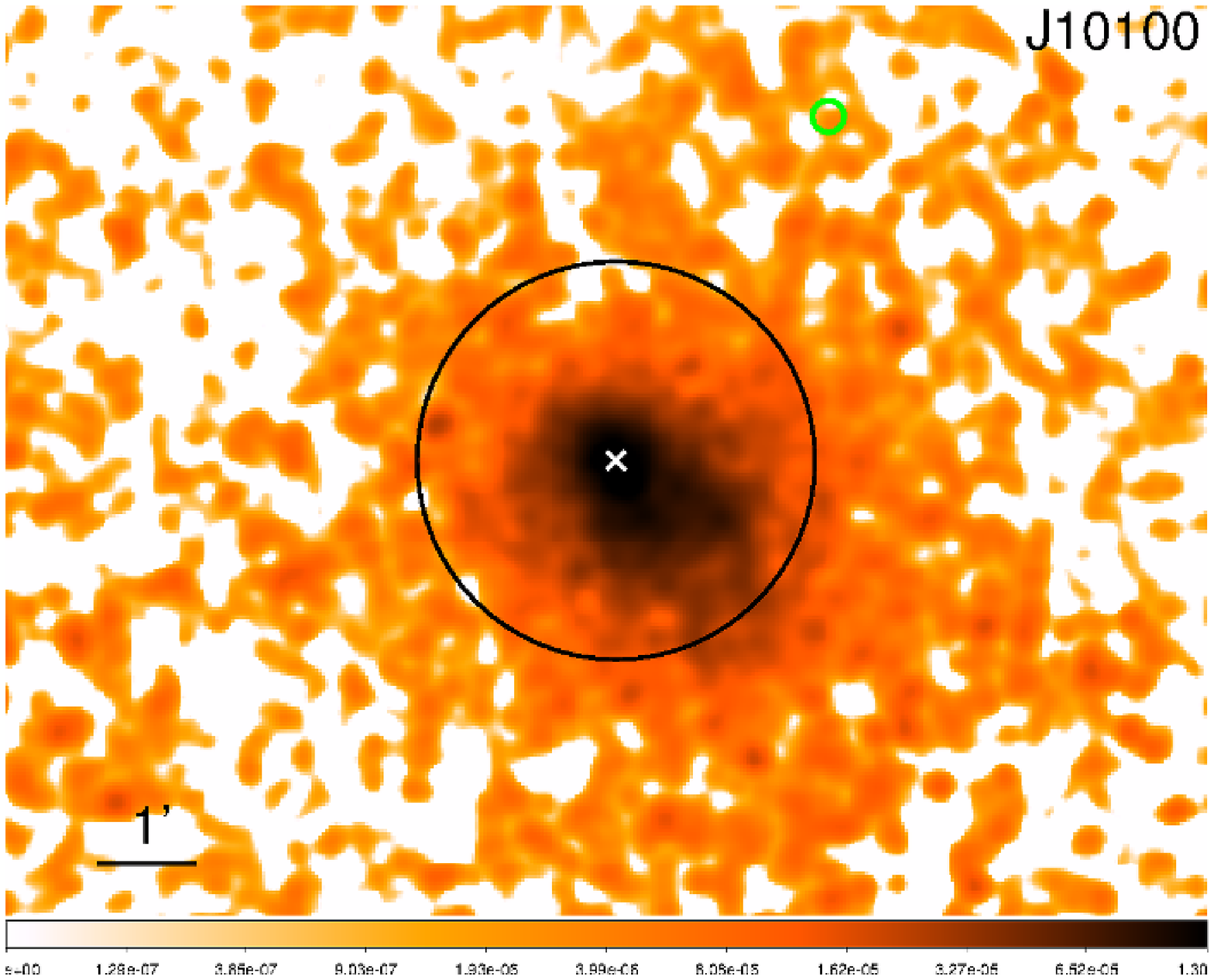} \vspace{0.2cm}
        \includegraphics[width=0.4\textwidth]{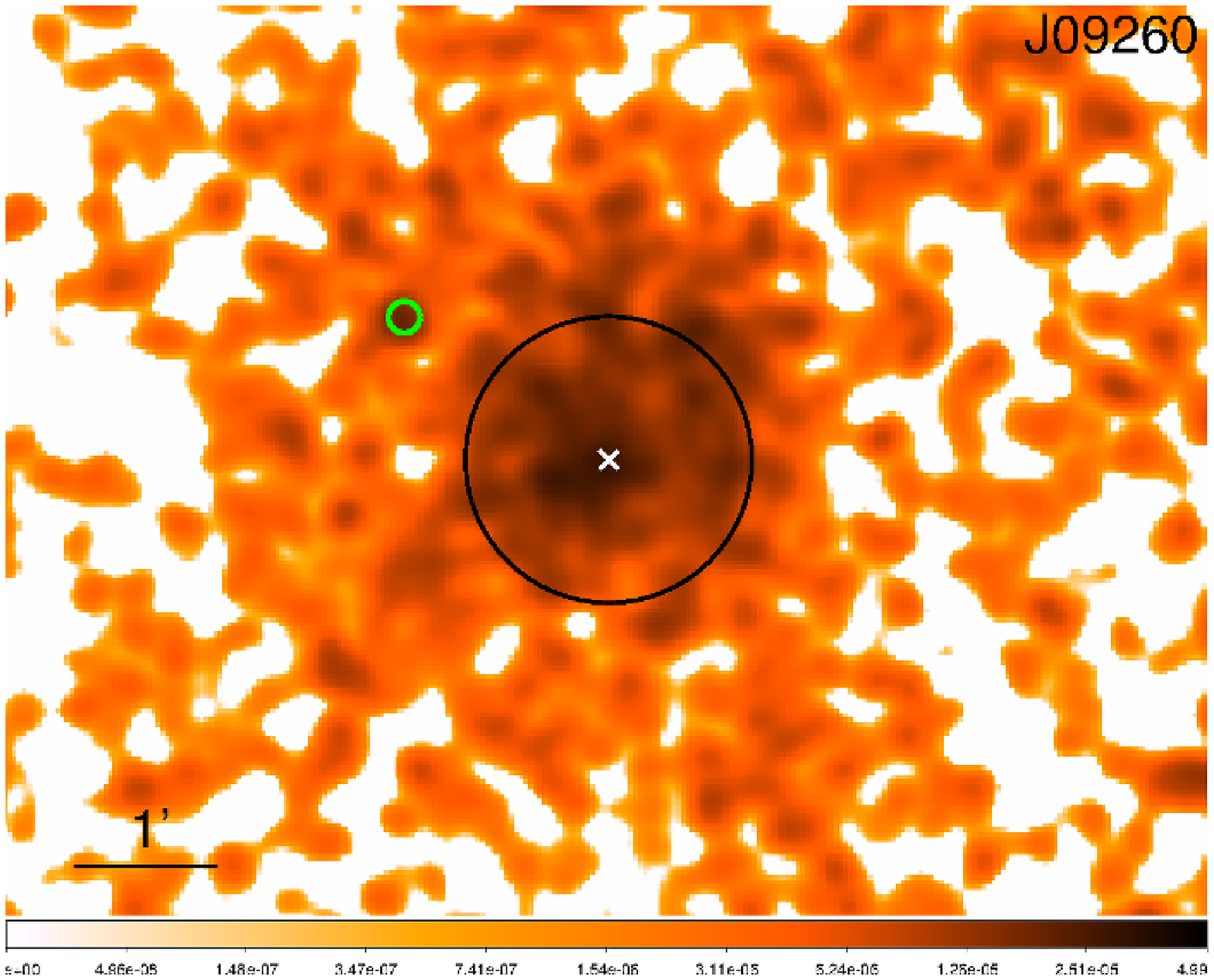} \hspace{0.2cm} \vspace{0.2cm}
       \includegraphics[width=0.4\textwidth]{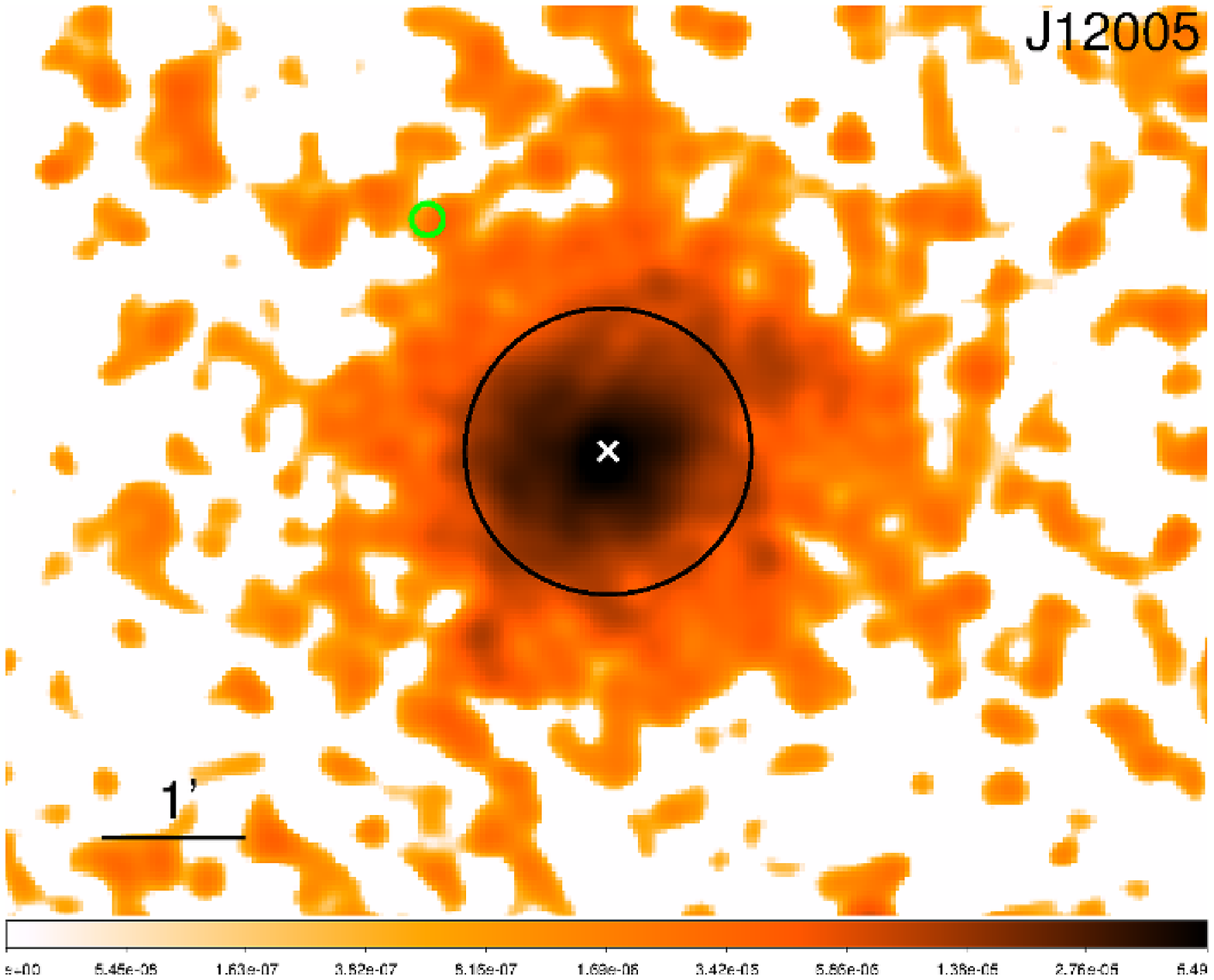} \vspace{0.2cm}
       \includegraphics[width=0.4\textwidth]{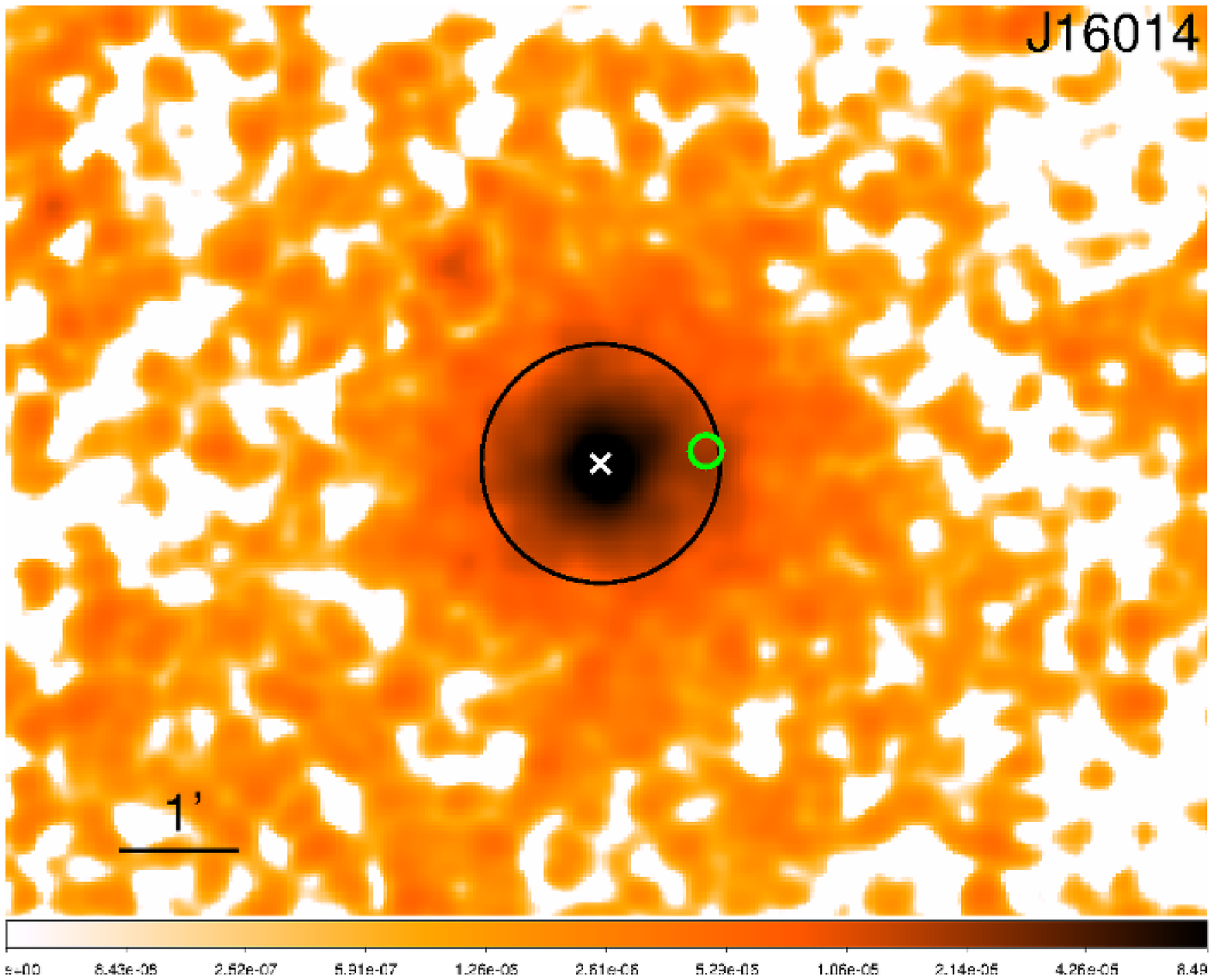}
\caption{\small{$0.3-2$ keV band \textit{XMM-Newton} X-ray images of the galaxy clusters. The data from EPIC MOS and PN cameras are combined. All background components are subtracted and exposure correction is applied. On every image, the large circular region (black) has a radius of $1\arcmin$ and is centered on the center of the galaxy cluster, where the center of the cluster is marked with the cross (white). The small circles (green) show the positions of the candidate BCGs. Only J10405 resides at the center of its host galaxy cluster with a separation of  $\approx4''$ (or $\approx15.5$ kpc). This suggests that an actively star-forming galaxy resides at the center of an $L_{\rm 500} = 2.7\times10^{43} \ \rm{erg \ s^{-1}}$ galaxy cluster. The other superluminous disk galaxies are located at large projected radii, $150-1070$~kpc, from the centers of the galaxy clusters.}}
\label{fig:halo2}
  \end{center}
\end{figure*}

\section{Results}
\label{sec:results}

\subsection{Images}
\label{sec:images}

In Figure \ref{fig:candidates}, we present the SDSS $g$-band images of the five superluminous disk galaxies that are candidate BCGs. Visual inspection shows that J10405 has a symmetric morphology and is surrounded by numerous low-mass satellite galaxies. The bulge-disk decomposition reveals the bulge dominated nature of this galaxy with a bulge fraction of $B/T = 0.66$ \citep{simard11}. Three galaxies,  J10100, J09260, and J12005, have small bulges ($B/T \lesssim 0.20$) and prominent spiral arms. The galaxy, J16014, resides in a dense environment with several satellite galaxies in its proximity. This galaxy has two bulges and its morphology suggests that it is undergoing a merger. 

We derive the stellar mass of the galaxies using FAST (Fitting and Assessment of Synthetic Templates) code \citep{2009ApJ...700..221K}. We utilize multi-band SDSS photometric data (\textsl{u, g, r, i, z}) and use the  redshift of the galaxies as input. Moreover, we assumed solar metallicity, a Salpeter initial mass function \citep{salpeter55}, and a Milky Way dust law for the extinction \citep{cardelli89}. The  stellar mass of the galaxies is in the range of $(1.2-5.2) \times 10^{11} \ \rm{M_{\odot}}$, which is about an order of magnitude higher than the typical stellar mass of spiral galaxies \citep{kelvin14}, and is comparable with the stellar mass of the most massive spirals in the local universe \citep{bogdan13a}. Based on the \textit{WISE} $12\ \rm{\mu m}$ monochromatic luminosities, \citet{ogle16} derived the star  formation rates of these galaxies as $(9.3 -28.2) \ \rm{M_{\odot} \ yr^{-1}} $, which demonstrates that they are actively star-forming, unlike typical BCGs. 

In Figure \ref{fig:halo2}, we present the background subtracted $0.3-2$ keV band \textit{XMM-Newton} X-ray images of the galaxy clusters. To maximize the signal-to-noise ratios, we merged the EPIC MOS and PN data. The images reveal detection of each galaxy cluster in our sample. The galaxy cluster in the proximity of J10100 exhibits a disturbed morphology, suggesting that it may be undergoing a merger. All other galaxy clusters have  symmetric morphologies and appear to be relaxed. 

In the X-ray images of the galaxy clusters, we mark the cluster center (see Section \ref{sec:center}) as well as the position of the candidate BCGs. This comparison reveals that only J10405 is coincident with the cluster center, while all other galaxies are offset from the center.

\begin{figure*}[!]
  \begin{center}
    \leavevmode
    \hspace{-0.2cm}
      \epsfxsize=0.34\textwidth \epsfbox{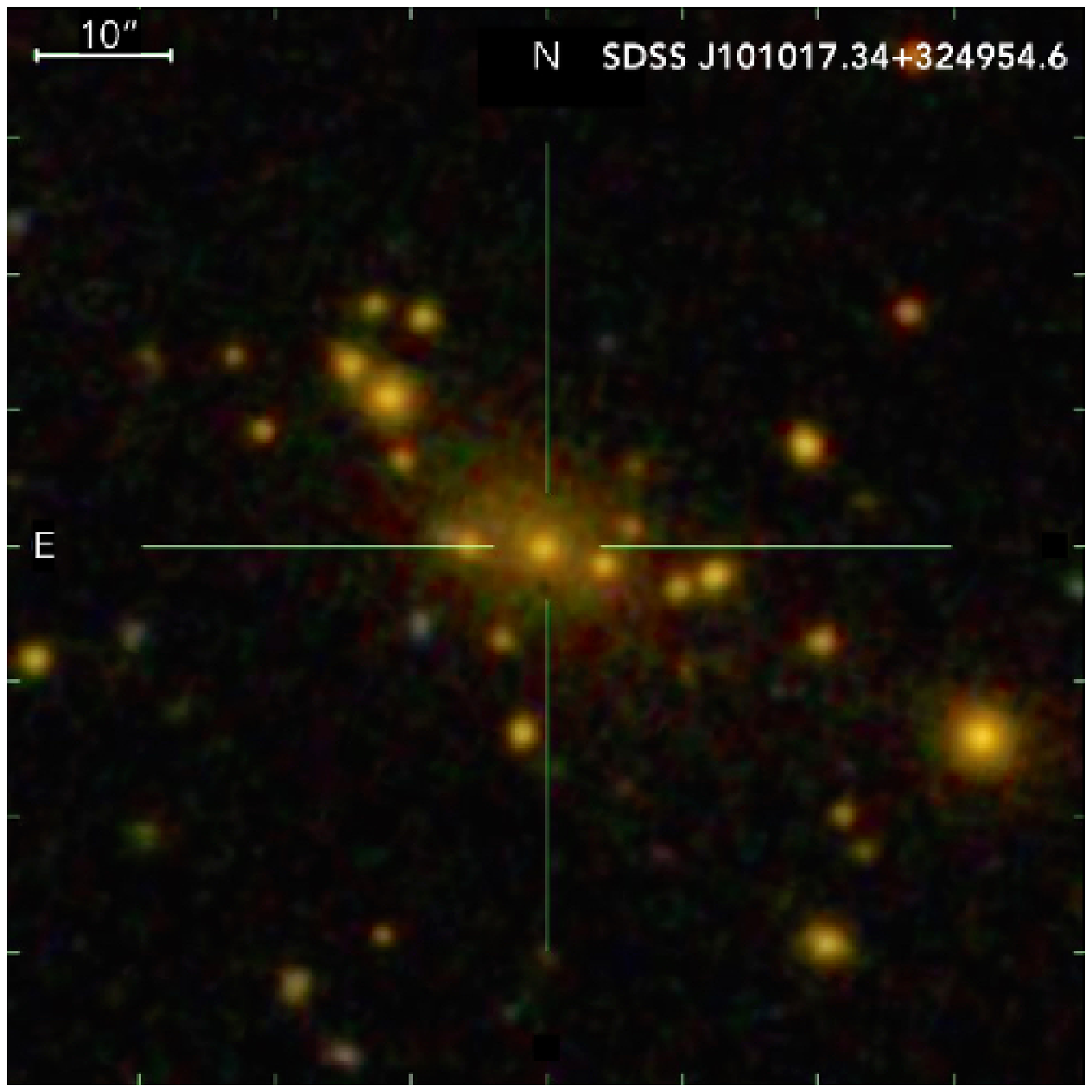}
      \vspace{0.1cm}
      \epsfxsize=0.34\textwidth \epsfbox{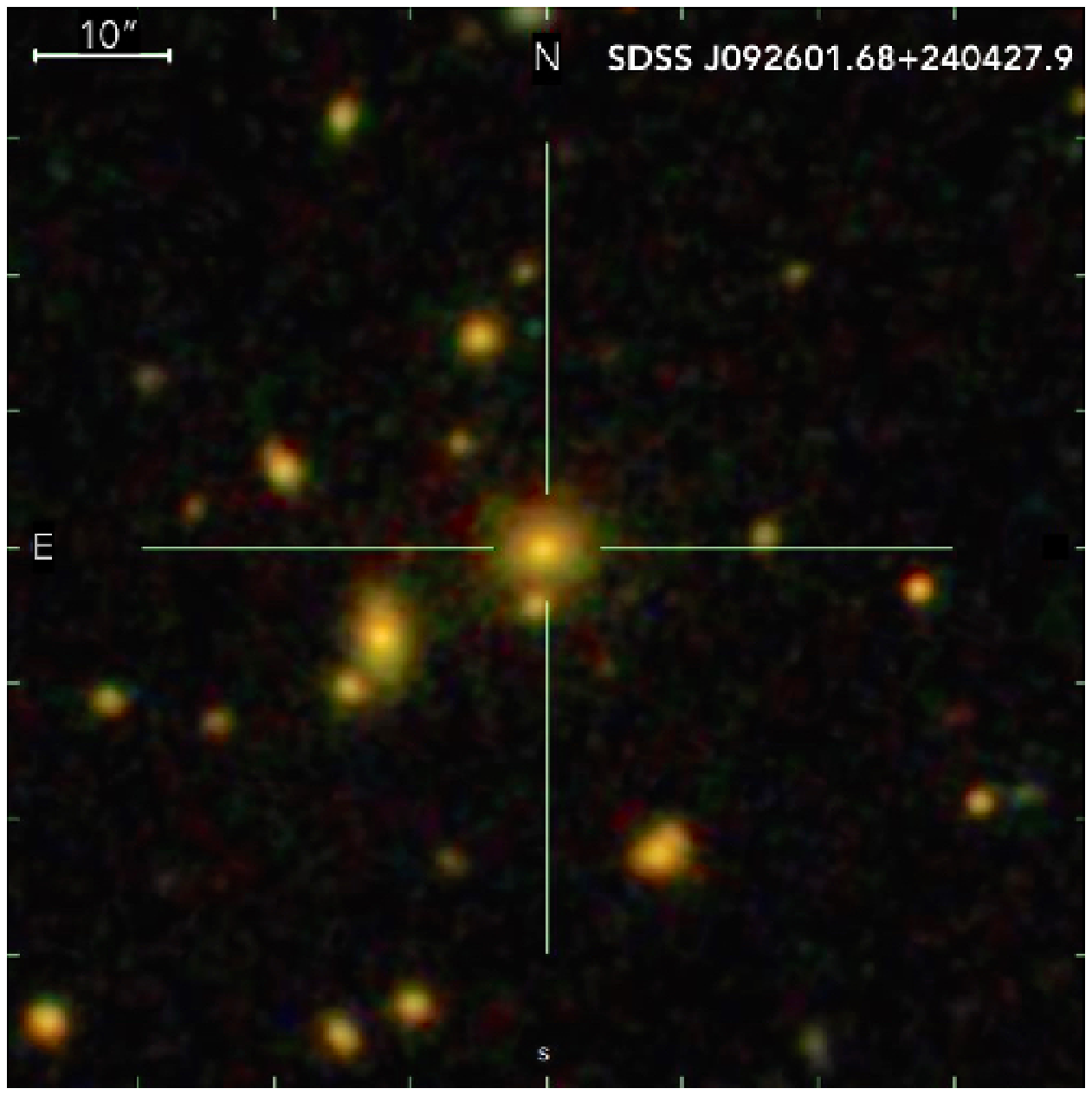}
      \epsfxsize=0.34\textwidth \epsfbox{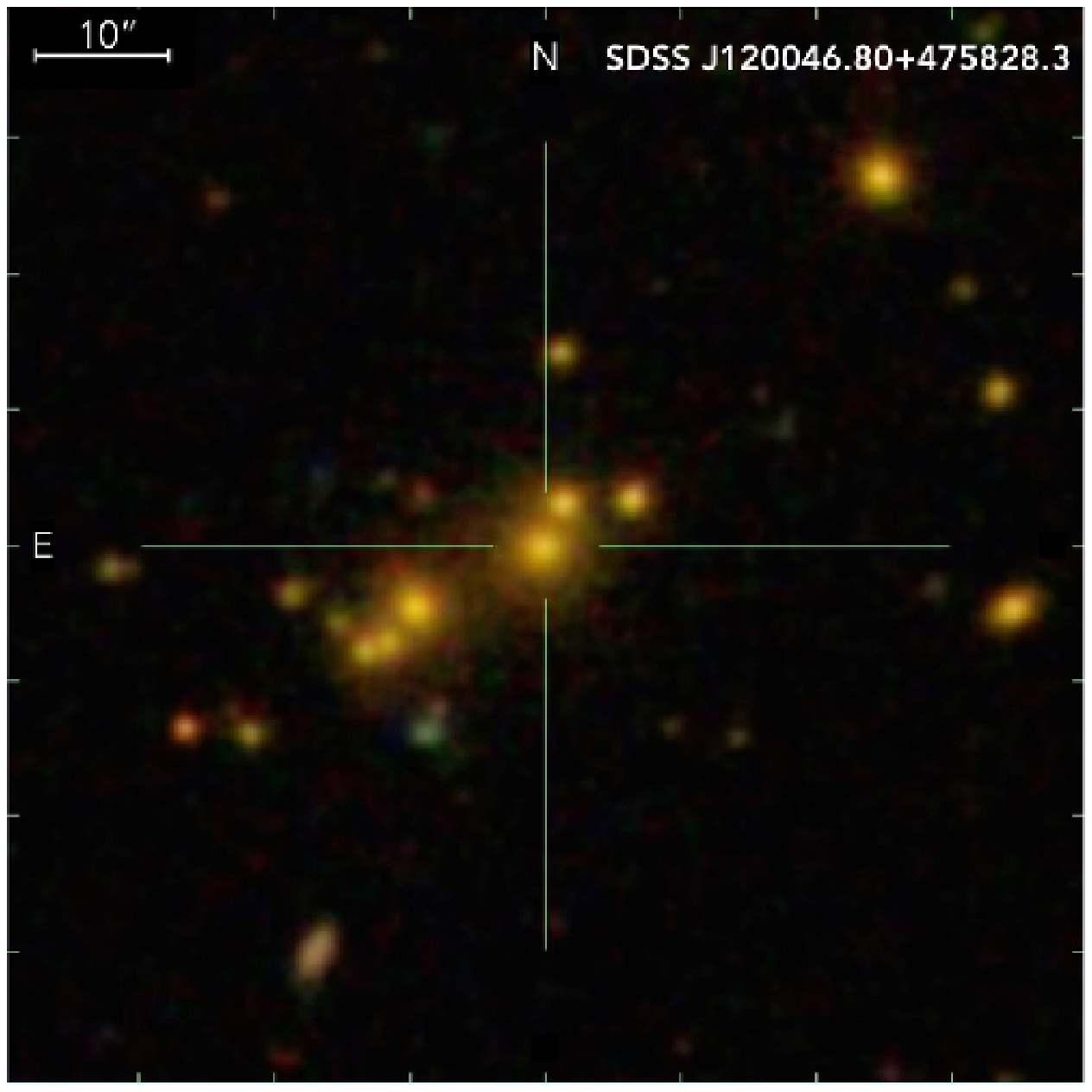}
      \epsfxsize=0.34\textwidth \epsfbox{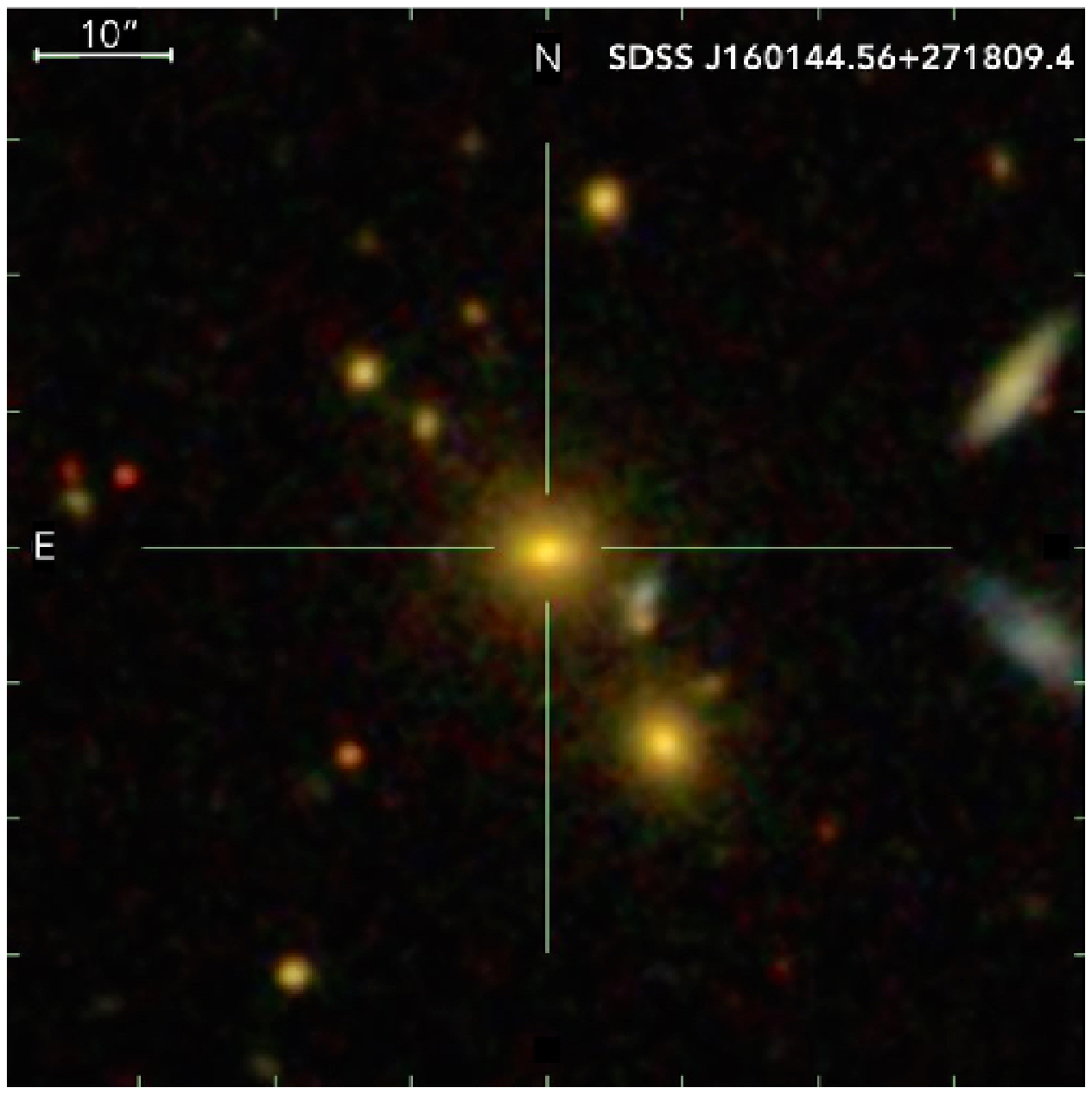}
      \vspace{0cm}
      \caption{SDSS $g$-band images of the central regions of the four galaxy clusters, whose superluminous spirals lie far from the cluster center. The galaxies that reside in the center of the clusters (shown with the cross-hairs) are quiescent, red ($u-r = 3.40-4.53$),  massive ($M_{\rm \star} = (1.9-8.5) \times 10^{11} \ \rm{M_{\odot}}$) elliptical galaxies. } 
     \label{fig:realcent}
  \end{center}
\end{figure*}

\subsection{Finding the cluster centers}
\label{sec:center}

To further investigate the central regions of the clusters, we determine the galaxy cluster center following the procedure outlined in \citet{lovisari15}. We smooth the $0.3-2$ keV band images  with a Gaussian with a kernel size of 3 pixels ($\approx12\arcsec$). On the smoothed image we search for the peak of the emission, which defines the cluster center. 

The offset between the position of J10405 and the peak of the X-ray emission is $\approx4\arcsec$, which corresponds to a projected distance of $15.5$ kpc. This offset is typical between the X-ray peak and BCG position \citep{zhang11}. The SDSS image demonstrates that J10405 is the most luminous galaxy in the central regions of the cluster (Figure \ref{fig:candidates}). While it would be interesting to probe the velocity difference between J10405 and the galaxy cluster, the redshift of the host cluster is not based on spectroscopic measurement \citep{ogle16}. As such, this redshift is not sufficiently accurate to infer the peculiar velocity of the galaxy.  Based on these, we conclude that the superluminous lenticular galaxy, J10405, resides at the bottom of the potential well and is the central BCG of a galaxy cluster. 

In principle, it is possible that the small separation between the positions of the X-ray peak and the BCG is caused by chance coincidence. To estimate the likelihood of chance coincidence, we utilized Monte Carlo simulations. The galaxy cluster hosting J10405 has 8 member galaxies within 10~Mpc \citep{ogle16}. To perform the simulations, we assumed that there are 8 galaxies within a 10~Mpc region and measured the likelihood that a randomly positioned galaxy  has a separation of $<15.5$~kpc. We repeated this simulation $10^6$ times to obtain statistically meaningful results. We found that the likelihood of randomly assigning a galaxy with a separation of $<15.5$ kpc is $\sim0.012$.

For all other systems in our sample, there is a large offset between the center of the galaxy cluster and the coordinates of the superluminous disks. Specifically, the offsets between the position of the galaxy and the cluster center are in the range of $150-1070$ kpc, which correspond to $(0.27-1.18)R_{\rm 500}$ (Table \ref{tab:clusters}). These large projected distances exceed the typical offsets between BCGs and the centers of X-ray--selected clusters \citep{zhang11}, thereby excluding the possibility that superluminous disks reside at the bottom of the cluster's potential well. However, given that the redshift of the candidate superluminous disk galaxies and the clusters are comparable, it is likely that these galaxies are members of the clusters.

\subsection{Characterizing the clusters}
\label{sec:spectra}

To characterize the galaxy clusters, we first determine the best-fit ICM temperatures within the $R_{\rm 500}$ radius. We derive the $R_{\rm 500}$ radius using an iterative process. Specifically, we compute the signal-to-noise ratio in the $0.3-2$ keV band as a function of radius using concentric annuli centered on the cluster center. We extract the initial spectrum within the region with the highest signal-to-noise ratio. We fit this spectrum and infer the initial temperature of the cluster. Using this initial  temperature and the best-fit $kT-R_{\rm 500}$ relation of \citet{arnaud05}, we infer the $R_{\rm 500}$ radius of the cluster. Then, we derive the best-fit ICM temperature within the new $R_{\rm 500}$ and utilize again the $kT-R_{\rm 500}$ relation to compute a new $R_{\rm 500}$ for the next iteration. We continue this process until the best-fit temperature remains invariant within $5\%$. 

To fit the ICM spectrum, we used an absorbed optically-thin thermal plasma emission model (\textsc{apec} in \textsc{XSpec}). We fixed the column density at the Galactic value for all galaxy clusters \citep{kalberla05}.  We allowed the  metal abundances to vary and used the abundance table of \citet{asplund09}. The temperature and normalization of the spectra were also free parameters. Because \textit{XMM-Newton} has three EPIC cameras, we extracted spectra from each of the cameras, whose spectra were fit simultaneously. Since some observations are still affected by soft proton contamination (see Section \ref{sec:data})  we added an extra power-law, folded only with the Response Matrix File, to the background modeling to account for that component.
  
The fitting procedure yields the ICM temperature and, through the normalization of the \textsc{apec} model, the $L_{\rm 500}$ luminosities (Table \ref{tab:clusters}). The temperatures and luminosities are in the range of $kT = 1.8-4.7 $ keV and $L_{\rm 500} = (0.3-3.7)\times10^{44} \ \rm{erg \ s^{-1}}$, signifying that all object in our sample are  galaxy clusters.

At the center of the galaxy cluster, SDSS CE J160.241898Ð01.069106, resides the superluminous lenticular galaxy, J10405. This galaxy cluster has an ICM temperature of $kT = 1.90\pm 0.40$~keV and a luminosity of  $L_{\rm 500} = 2.7\times10^{43} \ \rm{erg \ s^{-1}}$, which implies a total mass of $M_{\rm 500} = 10^{14} \ \rm{M_{\odot}} $. To verify the X-ray luminosity of the galaxy cluster, we also estimated its luminosity within $R_{\rm 500}$ from the count rates observed on EPIC PN. Based on the count rate of $C = (3.2\pm0.4) \times10^{-2} \ \rm{s^{-1}}$, we estimate a luminosity of $L_{\rm 500} = (1.4\pm0.2)\times10^{43} \ \rm{erg \ s^{-1}}$, which is lower than that obtained from the spectral fitting. According to the $L_{\rm 500} - kT$ relation \citep{arnaud05}, this lower luminosity would correspond to a $kT \sim 1.7$~keV galaxy cluster. Within statistical uncertainties, this inferred temperature is in agreement with that obtained from the \textsc{XSpec} fitting, but the lower luminosity and temperature may suggest that SDSS CE J160.241898Ð01.069106 is a massive galaxy group rather than a galaxy cluster.

\section{Discussion}
\label{sec:discussion}

\subsection{Galaxies at the cluster centers}
\label{sec:realcent}

We established that for four galaxy clusters in our sample, the superluminous spiral galaxy is not coincident with the peak of the X-ray emission. This raises the question about the nature and characteristics of the galaxies at the center of the potential well. Therefore, we identified the optically most luminous galaxy close to the X-ray peak by using the coordinates of the cluster center (Section \ref{sec:center}) and SDSS images of the central regions of the cluster. For each cluster, we could find a massive galaxy with projected separations of $1.2-32.8$~kpc with the mean of $16.6$~kpc.

The SDSS $g$-band images of the galaxies at the cluster centers are shown in Figure \ref{fig:realcent}. Each galaxy resides in dense galaxy environments. The optical appearance of these galaxies are symmetric and do not show signatures of star-forming disks. Their $u-r$ color index is in the range of $u-r = 3.40-4.53$, which hints that they are quiescent early-type galaxies \citep{strateva01}. The stellar mass of the galaxies is $M_{\rm \star} = (1.9-8.5) \times 10^{11} \ \rm{M_{\odot}}$, which values are typical for massive elliptical galaxies. Thus, the galaxies at the center of four clusters in our sample are giant elliptical galaxies.

Although the superluminous spiral galaxies do not reside in the center of four galaxy clusters, these galaxies are brighter than the elliptical galaxies at the bottom of the potential well. Specifically, the superluminous spirals are $0.34-1.38$ magnitudes brighter in the SDSS $g$-band than the central ellipticals. Therefore, the superluminous spiral galaxies are the true BCGs of these clusters despite the fact that they are at large projected distance from the cluster center.

 \subsection{Formation of superluminous disk galaxies at cluster centers}
  
Although intense star formation is detected in the cores of several galaxy clusters, the star formation in these systems is associated with cooling flows and not with the disk of the BCG  \citep{mcnamara07,mcdonald12}. Moreover, cooling flow clusters host giant ellipticals at their centers and not galaxies with extended disks. 

Throughout their evolution, galaxies undergo a series of mergers, which have a direct influence on the morphology of galaxies. Major mergers are believed to play an essential role in transforming disk galaxies to ellipticals \citep[e.g.][]{barnes92}. Moreover, minor mergers can also destroy the disks of galaxies \citep[e.g.][]{quinn86,velazquez99}. Therefore, in dense environments with frequent galaxy-galaxy interactions elliptical galaxies are the dominant population, while in low density environments disk galaxies are more numerous \citep[e.g.][]{dressler80}. Therefore, the detection of a superluminous lenticular at the center of a galaxy cluster is unexpected when considering the typical evolutionary path of galaxies. 

Simulations suggest that the merger of gas rich galaxies could result in the formation of an outer stellar disk \citep[e.g.][]{barnes02,springel05,robertson06}. In this picture, galaxy disks can re-form after a merger event. Recently, \citet{zhu18} utilized the IllustrisTNG simulation and  investigated how extended disk galaxies may form after mergers. Specifically, they identified a disk galaxy in the simulation with similar characteristics to Malin 1, which is an extremely extended disk galaxy in the local universe \citep{bothun87}. While Malin 1 is more than an order of magnitude less massive than J10405, its disk has a similar size to that of J10405. By tracing the evolution of the galaxy, \citet{zhu18} concluded that the large gas disk originated from the cooling of hot halo gas, which formed during the merger of a pair of galaxies. We note that this simulated galaxy resides in relative isolation and not in a galaxy cluster. While the IllustrisTNG volume does not contain similar objects in the core of clusters, it is likely that the frequency of such objects is too low to be observed in a simulation with relatively small volume.

While the star-formation rate, disk size, and stellar mass of J10405 are similar to other superluminous disk galaxies, the bulge-to-disk decomposition suggests the more bulge-dominated nature of this galaxy \citep{simard11}. Indeed, J10405, exhibits the highest bulge-to-disk ratio with $B/T = 0.66$ in the sample of 53 superluminous disk galaxies \citep{ogle16}.  This value is significantly higher than the median value of $B/T = 0.17$ obtained for the superluminous disk galaxies. Moreover, similar $B/T$ ratios are typically observed for S0/a and Sa galaxies \citep{graham08}. Considering the above described evolutionary scenario, it is possible that the progenitor of J10405 was a massive elliptical galaxy, which was identical with the current bulge of J10405. During a merger event the galaxy disk was re-formed, which resulted in the presently observed superluminous lenticular galaxy. Therefore, the characteristics of J10405 and results of the simulations argue that the disk of J10405 was not retained throughout the evolution of the galaxy, but it re-formed during merger activity. As a caveat, we note that J10405 is possibly an outlier in the sample of superluminous disk galaxies given its bulge-dominated nature. Therefore, it is feasible that other superluminous spirals with lower bulge-to-disk ratios, especially those residing in isolated environments, may have evolved differently. 

\smallskip

\begin{small}
\noindent
\textit{Acknowledgements.}
We thank the referee for the constructive report, which greatly improved the manuscript. We thank Patrick Ogle for his helpful comments. This work uses observations obtained with \textit{XMM-Newton}, an ESA science mission with instruments and contributions directly funded by ESA Member States and NASA. In this work, the NASA/IPAC Extragalactic Database (NED) have been used. \'A.B., C.J., W.R.F, and R.K. acknowledges support from the Smithsonian Institution. 
\end{small}

\bibliographystyle{apj}
\bibliography{paper2.bib} 

\end{document}